\documentclass[reprint,aps,prl,showkeys,groupedaddress,longbibliography,floatfix]{revtex4-1}
\usepackage{graphicx}        
\usepackage{grffile}         
\usepackage{graphics}
\usepackage{epsfig}
\usepackage{verbatim}
\usepackage{dcolumn}        
\usepackage{bm}            
\usepackage{amsmath}
\usepackage{amssymb}
\usepackage{mathtools}
\usepackage{physics}
\usepackage{hyperref}        
\hypersetup{colorlinks,breaklinks,
            linkcolor=black,urlcolor=black,
            anchorcolor=black,citecolor=black}
\usepackage{color}
\usepackage{float}
\newcommand{\EQ}[1]{Eq.~(\ref{eq:#1})}
\newcommand{\EQS}[2]{Eqs.~(\ref{eq:#1}) and (\ref{eq:#2})}
\newcommand{\EQSrange}[2]{Eqs.~(\ref{eq:#1} - \ref{eq:#2})}
\newcommand{\FIG}[1]{Fig.~\ref{fig:#1}}

\def\bal#1\eal{\begin{align}#1\end{align}}

\usepackage[dvipsnames]{xcolor}
\usepackage{xspace}

\begin{document}
\title{Optimal transport and anomalous thermal relaxations}

\author{Matthew R. Walker}
\affiliation{Department of Physics, University of Virginia, Charlottesville, VA 22904, USA}
\author{Saikat Bera}
\affiliation{Department of Physics, University of Virginia, Charlottesville, VA 22904, USA}
\author{Marija Vucelja}
\email{mvucelja@virginia.edu}
\affiliation{Department of Physics, University of Virginia, Charlottesville, VA 22904, USA}
\affiliation{Department of Mathematics, University of Virginia, Charlottesville, VA 22904, USA}

\begin{abstract}
We study connections between optimal transport and anomalous thermal relaxations. A prime example of anomalous thermal relaxations is the Mpemba effect, which occurs when a hot system overtakes an identical warm system and cools down faster. Conversely, optimal transport is a resource-efficient way to transport the source distribution to a target distribution in a finite time. By "a resource-efficient way," what is often meant is with the least amount of entropy production. Our paradigm for a continuum system is a particle diffusing on a potential landscape, while for a discrete system, we use a three-state Markov jump process. In the continuous case, the Mpemba effect is generically associated with high entropy production. As such, at large yet finite times, the system evolution toward the target is not optimal in this respect. However, in the discrete case, we show that for specific dynamics, the optimal transport and the strong variant of the Mpemba effect can occur for the same relaxation protocol. 
\end{abstract}

\keywords{Thermal relaxation, Mpemba effect, Thermal quench, Optimal transport, Wasserstein distance, Minimal Flow cost}

\maketitle
The Mpemba effect represents a "shortcut" to equilibrium. It is thus natural to ask if this shortcut to equilibrium is related to a well-known problem of optimal transport. Here we describe a relation between the two in the case of Markov jump processes. 

Optimal transport is a rich mathematics and statistics problem concerned with the optimal way of transporting a distribution from a source to a target function in a finite amount of time. The problem has a long history, starting with Monge 1781, who formalized it and illustrated with an example of the most economical way of transporting soil from one place to another~\cite{monge_memoire_1781}. Major advances and connections to linear programming were later made by Kantorovich~\cite{kantorovich_translocation_1942,villani_optimal_2009}. The applications of the specific solution to the optimal problem span a variety of fields, such as, e.g., statistics and machine learning~\cite{kolouri_optimal_2017}, molecular biology~\cite{schiebinger_optimal-transport_2019}, classical mechanics~\cite{koehl_statistical_2019}, linguistics~\cite{huang_supervised_2016} and computer vision~\cite{haker_optimal_2004}. Recently geometrical~\cite{chennakesavalu_unified_2023}, thermodynamical~\cite{van_vu_thermodynamic_2023}, and topological~\cite{van_vu_topological_2023} interpretations of the aspects of the optimal transport problem were made. The thermodynamical interpretation is especially relevant in stochastic thermodynamics~\cite{seifert_stochastic_2012}.

Besides optimal routes to a target distribution, fast routes are also of interest. One such "shortcut" is the Mpemba effect -- a counter-intuitive relaxation process in which a system starting at a hot temperature cools down faster than an identical system starting at an initially lower temperature when both are coupled to an even colder bath. An analogous effect exists in heating. By now Mpemba effect was seen in water~\cite{mpemba_cool_1969}, colloidal systems~\cite{kumar_exponentially_2020,kumar_anomalous_2022}, polymers~\cite{hu_conformation_2018}, magnetic alloys~\cite{chaddah_overtaking_2010}, clathrate hydrates~\cite{ahn_experimental_2016}, granular fluids~\cite{torrente_large_2019,lasanta_when_2017}, spin glasses~\cite{baity-jesi_mpemba_2019}, quantum systems~\cite{nava_lindblad_2019}, nanotube resonators~\cite{greaney_mpemba-like_2011}, cold gasses~\cite{keller_quenches_2018}, mean-field antiferromagnets~\cite{klich_mpemba_2019}, systems without equipartition~\cite{gijon_paths_2019}, molecular dynamics of water molecules~\cite{jin_mechanisms_2015}, driven granular gasses~\cite{biswas_mpemba_2020}, and molecular gasses~\cite{santos_mpemba_2020}. The Mpemba effect was formulated for a general Markovian system in~\cite{lu_nonequilibrium_2017}. The strong variant of the effect, the so-called Strong Mpemba effect, was introduced in~\cite{klich_mpemba_2019} and experimentally observed in~\cite{kumar_exponentially_2020}. Optimal heating strategy applications were discussed in~\cite{gal_precooling_2020}. The Mpemba effect in the overdamped limit of a particle diffusing on a potential landscape was studied in~\cite{lu_nonequilibrium_2017,walker_anomalous_2021,walker_mpemba_2023,chetrite_metastable_2021}. Other theoretical advances involving the Mpemba phenomenon link it to phase transitions~\cite{holtzman_landau_2022}, relaxations to nonequilibrium steady states~\cite{degunther_anomalous_2022}, Otto cycle efficiency~\cite{lin_power_2022}, stochastic resetting~\cite{busiello_inducing_2021}, random energy models~\cite{klich_mpemba_2019}, quantum analogs in Lindblad dynamics~\cite{nava_lindblad_2019,carollo_exponentially_2021,chatterjee_quantum_2023}, and quantum analogs related to symmetry breaking~\cite{ares_entanglement_2023}. Recently the effects of the type of coupling between the system and the bath~\cite{teza_far_2022,teza_relaxation_2021}, effects of dynamics~\cite{bera_effect_2023} and of eigenvalue crossings~\cite{teza_eigenvalue_2023} on the phenomenon were studied. 

The Mpemba effect can be viewed as an optimization of the initial condition. However, in scenarios where the initial conditions are fixed, sometimes we can vary the dynamics and obtain an analogous effect starting from the same initial condition but with different dynamics~\cite{bera_effect_2023}. Below we refer to this similar effect as the Mpemba effect. In that case, our initial and final points in the probability distribution space are fixed, and the problem starts to resemble a problem of transport from a source to a target. Here we ask if there are cases in which the same dynamics corresponds to the optimal transport, i.e., minimal entropy production and the Mpemba effect. Our enabling examples are the two main paradigms of stochastic thermodynamics -- a particle diffusing over a potential landscape with overdamped-Langevin dynamics and a Markov jump process. 

Surprisingly, the strong variant of the Mpemba effect in certain discrete cases coincides with the optimal transport.
Below we show that the results depend on the large time we are looking at, the relaxation modes, and net probability currents. 

The paper is organized as follows. We first introduce the notation relevant to Markov jump processes. Next, we present the optimal transport and the Wasserstein distance as a good measure of optimal transport. We continue by introducing the Mpemba effect. Afterward, we discuss anomalous thermal relaxations and optimal transport for a particle diffusing on a potential landscape and a three-state Markov jump process. We finish with a discussion of the results. 

\section{Setup and notation}
Although we consider continuous and discrete examples, it is instructive to introduce first the concepts and notations of entropy production, mobility, and the Mpemba effect on Markov jump processes. 

We consider a Markov jump process, which obeys the Master equation
\begin{eqnarray}
\label{eq:Mastereq}
    \partial _t p = R\,p,  
\end{eqnarray}
where $p_x(t)$ is the probability of finding the system is state $x\in\Omega$ at time $t$, and $R$ is the rate matrix, with $R_{xy}$ as the transition rate from $y$ to $x$.  Each state $x$ is characterized by energy $E_x$. We consider rate matrices that obey Detailed Balance (DB), 
\begin{align}
\label{eq:detailedbalance}
    R_{xy}\pi^{T_b}_y = R_{yx}\pi^{T_b} _x. 
\end{align}
where $\pi^{T_b}$ is stationary solution of~\EQ{Mastereq} system -- the Boltzmann distribution
\begin{eqnarray}
    \pi _{T_b,x} = \frac{1}{Z({T_b})}e^{-\beta_bE_x}, 
\end{eqnarray}
with $Z({T_b}) = \sum_{x \in \Omega}\exp\left[-\beta_bE_x\right]$ as the partition sum. Below we label $\beta_b = 1/(k_BT_b)$ and set the Boltzmann constant to be unity, $k_B = 1$. 

It is useful to define the following quantities. The entropy change after a state change from $y$ to $x$ is 
\begin{align}
\label{eq:entropychangeafterstatechange}
    s_{xy} = \ln \frac{R_{xy}}{R_{yx}} = \beta_b (E_y - E_x). 
\end{align}
The frequency of jumps from $y$ to $x$ at $t$ is 
\begin{align}
\label{eq:jumpfreq}
    a_{xy}(t) = R_{xy} p_y(t), 
\end{align}
and the probability current from state $y$ to $x$ at $t$ is 
\begin{align}
    j_{xy}(t) = R_{xy} p_y(t) - R_{yx} p_x(t). 
\end{align}

The \emph{dynamical activity} is the amplitude of the transitions between the states 
\begin{eqnarray}
    a(t)  = \sum _{\substack{x\neq y\\x,y\in\Omega}} a_{xy}(t). 
\end{eqnarray}
The average number of jumps during time $\tau$ is 
\begin{eqnarray}
    \mathcal{A}(\tau) = \int ^\tau _0 a(t)dt.  
\end{eqnarray}
The entropy of the system is the Shannon entropy
\begin{eqnarray}
    \label{eq:Shannonentropy}
    S(p) = - \sum _x p_x \ln p_x, 
\end{eqnarray}
thus the change in the entropy of the system is 
\begin{eqnarray}
    \Delta S_{\rm sys}= S(p(\tau)) - S(p(0)). 
\end{eqnarray}
The entropy change of the environment is 
\begin{eqnarray}
    \Delta S_{\rm env} = \int_{0}^{\tau} \sum _{\substack{x\neq y\\x,y\in\Omega}}a_{xy}(t)s_{xy}\, dt. 
\end{eqnarray}
By using ~\EQS{entropychangeafterstatechange}{jumpfreq}, it can be written in an explicit form as 
\begin{eqnarray}
    \label{eq:DeltaSenvexp}
    \Delta S_{\rm env}
= \displaystyle\sum _{x \in \Omega}\beta_bE_x\left[\pi ^T_x - p_x(\tau)\right].
\end{eqnarray}
The total entropy production is the sum of the change in the entropy of the environment and the change in the entropy of the system,
\begin{eqnarray} 
    \label{eq:totalentropyproductiondefinition}
   &\Sigma(\tau) = \Delta S_{\rm env}+ \Delta S_{\rm sys}.
\end{eqnarray}
Using~\EQSrange{Shannonentropy}{DeltaSenvexp}, the total entropy production is explicitly
\begin{eqnarray}
\nonumber
    \Sigma (\tau) &=& \displaystyle\sum _{x \in \Omega} \bigg\{ \beta_bE_x\left[\pi ^T_x - p_x(\tau)\right]
    \\
    \label{eq:totalentropyproduction}
    && + \pi _x^T \ln \pi _x ^T- p_x(\tau)\ln p_x(\tau) \bigg\}.
\end{eqnarray}
The entropy production rate, $\sigma (t)\equiv d\Sigma (t)/dt$, is 
\begin{eqnarray}
   \sigma (t) 
   &=& \sum _{\substack{x>y\\x,y\in\Omega}}\left(a_{xy}(t) - a_{yx}(t)\right) \ln \left[\frac{a_{xy}(t)}{a_{yx}(t)}\right],  
\end{eqnarray}
\cite{schnakenberg_network_1976,seifert_stochastic_2012}. Note that the entropy production rate is always non-negative, as $a_{xy} - a_{yx}$ and $\ln [a_{xy}/a_{yx}]$ always have matching signs. 

Close to equilibrium for macroscopic systems, the currents depend on the thermodynamic forces in a linear fashion. The coefficients of this linear dependence are the Onsager coefficients~\cite{onsager_reciprocal_1931,onsager_reciprocal_1931-1}. For microscopic systems far from equilibrium, one can define Onsager-like coefficients. The \emph{generalized force} between transitions is 
\begin{eqnarray}
    f_{xy}(t) = \ln \frac{a_{xy}(t)}{a_{yx}(t)}, 
\end{eqnarray}
see e.g.~\cite{van_vu_thermodynamic_2023}. The thermodynamic force is the sum of the entropy changes in the system and the environment. The ratio of the currents to the forces 
\begin{eqnarray}
\label{eq:Onsager-like-coefs}
    m_{xy}(t) = \frac{j_{xy}(t)}{f_{xy}(t)} = \frac{a_{xy}(t) - a_{yx}(t)}{\ln a_{xy}(t) - \ln a_{yx}(t)},
\end{eqnarray} 
identifies the "linear response" coefficients, $m_{xy}$, which play the microscopic analogs of the Onsager coefficients, as the entropy production rate can be expressed as a quadratic form of generalized forces
\begin{eqnarray}
    \sigma(t) = \sum _{\substack{x>y\\x,y\in\Omega}} m_{xy}(t) \left[f_{xy}(t)\right]^2. 
\end{eqnarray}
The sum linear response coefficients
\begin{eqnarray}
\label{eq:mobility}
m(t) = \sum _{\substack{x>y\\x,y\in\Omega}} m_{xy}(t), 
\end{eqnarray}
is the \emph{dynamical state mobility}, while the \emph{kinetic cost} is defined as 
\begin{eqnarray}
\label{eq:kineticcost}
    \mathcal{M}(\tau) = \int ^\tau _0 m(t)\, dt. 
\end{eqnarray}
For the overdamped-Langevin dynamics, the dynamical mobility converges to a constant proportional to the diffusion coefficient, $m \propto T_b$, and thus, the kinetic cost linearly scales with time,
\begin{eqnarray}
\label{eq:langevin-kinetic-cost}
    \mathcal{M}(\tau) \propto T_b \,\tau. 
\end{eqnarray}
Other introduced quantities have straightforward analogs in the continuous case. Next, we discuss the optimal transport solutions for continuous and discrete classical cases. 

\section{Optimal transport metric -- Wasserstein distance}
We look at cases where, given two probability distributions, at initial time $t=0$ and at the finite final time, $\tau$, and a protocol specifying the dynamics, there is an optimal transport protocol between the two, which minimizes the entropy production. The solution to the optimal transport problem provides an optimal transport plan between the source and target distributions. The \emph{Wasserstein distance} is a metric in the space of probability distributions useful in quantifying the optimality of the transport. Other names for this metric are the \emph{Monge-Kantorovich distance} or the \emph{earth mover's distance}. The Wasserstein metric was extensively studied, and has thermodynamics~\cite{van_vu_thermodynamic_2023}, geometric~\cite{chennakesavalu_unified_2023}, topological~\cite{van_vu_topological_2023}, and fluid mechanics~\cite{benamou_computational_2000,aurell_optimal_2011,aurell_refined_2012} interpretations. Below we define the Wasserstein distance and examine its meaning in the context of anomalous thermal relaxations. 

In the continuous case, there is a beautiful fluid mechanics interpretation of the optimal transport problem, given by Benamou and Brenier~\cite{benamou_computational_2000,aurell_optimal_2011,aurell_refined_2012}. Suppose the evolution of the probability density, $p(x,t)$, is governed by a continuity equation, 
\begin{eqnarray}
\label{eq:continuityeq}
    \partial _t p(x,t) + \partial_x [v(x,t) p(x,t)] = 0, 
\end{eqnarray}
then the $L_2-$Wasserstein distance from $p^A = p(0)$ at initial time $t=0$ to $p^B=p(\tau)$, at final time, $\tau$, is given by the so-called the Benamou-Brenier formula, 
\begin{eqnarray}
\label{eq:Benamou-Brenier-formula}
   \mathcal{W}_{2}(p^A, p^B) = \min _{v} \sqrt{T_b\tau \Sigma(\tau)},
\end{eqnarray}
where the total entropy production during period $\tau$ is 
\begin{eqnarray}
\label{eq:Ent-prod-continuous}
    \Sigma (\tau) = \frac{1}{T_b} \int^\tau _0 \int _\mathcal{D} \left[v(x,t)\right]^2 p(x,t) dx dt, 
\end{eqnarray}
~\cite{seifert_stochastic_2012}.
The Wasserstein distance $ \mathcal{W}_{2}(p^A, p^B)$ is minimum is over all smooth paths $\{ v(t)\}_{0 \leq t \leq \tau}$, subject to~\EQ{continuityeq}.  

For a discrete system evolving with the Master equation,~\EQ{Mastereq}, the  $L_1-$Wasserstein distance, $\mathcal{W}_1(p^A,p^B)$, is 
\begin{eqnarray}
    \mathcal{W}_1(p^A,p^B) = \displaystyle\min _{p^{AB}} \sum _{x,y\in \Omega} C_{xy}\, p ^{AB} _{xy}, 
\end{eqnarray}
where $C_{xy}$ is the cost function, $p^{AB} _{xy}$ is a joint distribution, with marginals corresponding to $p^A _x$ and $p^B _y$, and the minimum is taken over a set of all admissible couplings, see e.g.~\cite{kantorovich_translocation_1942,villani_optimal_2009}. The Wasserstein distance $\mathcal{W}_1(p^A,p^B)$ is bounded from above by
\begin{eqnarray}
    &\mathcal{W}_1&(p^A,p^B) \leq \mathcal{J}(\tau) 
    \leq  \mathcal{C}_{\sigma m}(\tau)\leq\mathcal{C}_{\Sigma \mathcal{M}}(\tau),
\end{eqnarray}
with $\mathcal{J}(\tau)$ as the flow cost, 
\begin{eqnarray}
\label{eq:flowcost}
    \mathcal{J}(\tau) &\equiv&  \int ^\tau _0 \sum _{\substack{x>y\\x,y\in\Omega}}|j_{xy}(t)|dt, 
\end{eqnarray}
upper bound $\mathcal{C}_{\sigma m}(\tau)$ which depends on entropy production rate and dynamical mobility
\begin{eqnarray}
    \mathcal{C}_{\sigma m}(\tau) &\equiv& \int ^\tau _0\sqrt {\sigma(t)m(t)}dt,  
\end{eqnarray}
and upper bound $\mathcal{C}_{\Sigma \mathcal{M}}(\tau)$ which depends on entropy production and kinetic cost, 
\begin{eqnarray}
    \mathcal{C}_{\Sigma \mathcal{M}}(\tau) &\equiv& \Sigma(\tau) \mathcal{M}(\tau),  
\end{eqnarray}
see~\cite{van_vu_thermodynamic_2023}. 
The Wasserstein distance is bounded below, 
\begin{eqnarray}
\label{eq:Wasserstein-totalvariation}
    \mathcal{W}_1\left[\pi^T, p(\tau)\right] \geq \mathcal{T} \left[\pi^T, p(\tau)\right],  
\end{eqnarray}
by the total variation distance $\mathcal{T}$, 
\begin{eqnarray}
\label{eq:totalvariationdistance}
    \mathcal{T} \left[\pi^T, p(\tau)\right]\equiv \frac{1}{2} \sum _{x\in\Omega} \left|\pi^T_x - p_x(\tau)\right|. 
\end{eqnarray}
Equality in~\EQ{Wasserstein-totalvariation} holds for the case of fully-connected graphs~\cite{van_vu_thermodynamic_2023}.

The following section introduces the Mpemba effect as an example of anomalous thermal relaxations. 

\section{Mpemba effect}
The Mpemba effect occurs when a system prepared at initial temperature $T_h$ and immersed in a bath of temperature $T_b$ relaxes faster down to the bath's temperature than a replica of the same system starting at $T_w$, where $T_b \leq T_w \leq T_h$,~\cite{lu_nonequilibrium_2017}. An analogous effect also occurs in heating, and it is called the inverse Mpemba effect~\cite{lu_nonequilibrium_2017}. 

Below we specify what we mean by the Mpemba effect on a classical discrete case, where the relaxation is governed by the Master equation~\EQ{Mastereq}. The generalization to continuous systems evolving with~\EQ{continuityeq}. Note that we restrict our considerations to systems with Markov property, i.e., the system's future state depends only on the present state. However, one can also consider systems with memory. Sometimes the Mpemba effect on systems with Markov property is called the Markovian Mpemba~\cite{lu_nonequilibrium_2017}. 

At large times a probability distribution of a relaxing system, evolving according to~\EQ{Mastereq}, that is initiated at temperature $T$, $p(0) = \pi^T$, is characterized by 
\begin{eqnarray}
    p(t) = \pi^{T_b} + \sum _{i > 1}a_i v_i e^{\lambda_i t}, 
\end{eqnarray}
where $\lambda_i$ are the eigenvalues of $R$, $v_i$ are the right eigenvectors of $R$, and $a_i$ are the overlap coefficients of the left eigenvector $u_i$ of the rate matrix $R$ and the initial condition, 
\begin{eqnarray}
    a_i \equiv \frac{\langle u_i, \pi^T\rangle}{\langle u_i,v_i\rangle}, 
\end{eqnarray}
The eigenvalues of $R$ are ordered and nonpositive, $\lambda _1 = 0 > \lambda_2 \geq \lambda_3 \geq ...$. We assume that there is a gap between $\lambda_2$ and $\lambda_3$, thus in the long time limit, the evolution of the system is 
\begin{eqnarray}
     p(t) \approx \pi^{T_b} + a_2 v_2 e^{\lambda_2 t}.  
\end{eqnarray}
The Mpemba effect occurs when the overlap coefficient  $a_2$ with respect to initial conditions is nonmonotonic~\cite{lu_nonequilibrium_2017}. That is if comparing two identical systems, prepared at $T_h$ and $T_w$, in their independent relaxation to thermal equilibrium at $T_b$, we have the Mpemba effect for $T_h \geq T_w \geq T_b$ and $|a_2(T_h)| \leq |a_2(T_w)|$. The Mpemba effect is the most pronounced if the slowest mode is orthogonal to the initial conditions, i.e., if $a_2(T_h) = 0$. In this case, the relaxation of the system approaches the equilibrium state from the direction of $u_3$, and there is a jump in the relaxation time from $-1/\lambda_2$ to $-1/\lambda_3$ at $T_h$. We refer to the case where there is no projection of the slow mode to the initial conditions as the Strong Mpemba effect. 

\subsection{Distance-from-equilibrium}

Distance-from-equilibrium should satisfy the following properties~\cite{lu_nonequilibrium_2017}: (i) during a relaxation process, the distance should monotonically decrease with time, (ii) the distance from a Boltzmann distribution at $T$ to equilibrium at $T_b$ is a monotonically increasing function of $|T - T_b|$, with, in general, different pre-factors for cooling and heating, and (iii) the distance is a continuous and convex function of $p(t)$. The suitable choices are, for example, the Kulback-Leibler divergence and $L_1$ norm,~\cite{lu_nonequilibrium_2017, chetrite_metastable_2021}. We define them below.  

The \emph{Kullback-Leibler} (KL) \emph{divergence}~\cite{kullback_information_1951}, is defined as 
\begin{eqnarray}
\label{eq:KLdiv}
   D_{\rm KL}\left(p(t)||\pi^{T_b}\right)\equiv \sum_{x \in \Omega} p_x(t) \ln \left[\frac{p_x(t)}{\pi_x^{T_b}}\right]. 
\end{eqnarray}
It can be thought of as the "entropic distance," by which we mean the total amount of entropy production in a relaxation process, starting from $p(t)$ and ending at $\pi^{T_b}$,  
\begin{eqnarray}
\int ^\infty _t \sigma (t')\,dt' = \Sigma (\infty) - \Sigma (t), 
\end{eqnarray}
see e.g.~\cite{lu_nonequilibrium_2017}. With~\EQ{totalentropyproduction}, the above expression can be written as
\begin{eqnarray}
\nonumber
\int ^\infty _t \sigma (t')\,dt'&=& \sum _{x \in \Omega} \bigg\{ \beta_bE_x\left[p_x(t)-\pi^{T_b}_x\right]
    \\
    &+&p_x(t)\ln p_x(t) - \pi _x^{T_b} \ln \pi _x ^{T_b} \bigg\}, 
\end{eqnarray}
which is the KL divergence,~\EQ{KLdiv}, hence 
\begin{eqnarray}
\label{eq:KLdivdiffSigma}
    D_{\rm KL}\left(p(t)||\pi^{T_b}\right) = \Sigma (\infty) - \Sigma (t).
\end{eqnarray}

The $L_1-$\emph{norm} is
\begin{eqnarray}
\left|\left|p(t),\pi^{T_b}\right|\right|_1 = \sum _{x\in \Omega} \left|p_x(t) - \pi^{T_b}_x\right|. 
\end{eqnarray}
Note that $L_1$-norm is twice the \emph{total variation distance} $\mathcal{T}$, see~\EQ{totalvariationdistance}.

Next, on examples of over-damped Langevin dynamics with metastability and a three-level system, we connect the concepts of optimal transport and the Mpemba effect. 

\section{Examples}

\subsection{Particle diffusion on a potential landscape}

Let us consider a Brownian particle subject to a potential force $- U' \equiv - dU/dx$ and suppose that the particle is subject to over-damped Langevin dynamics 
\begin{align}
\label{eq:overdampedLangevin}
    \gamma \frac{d}{dt} {\rm x}(t) = -\frac{1}{m}U'[{\rm x}(t)] + \Gamma(t), 
\end{align}
where ${\rm x}(t)$ is particle's trajectory, $\gamma$ is the friction coefficient, and $\Gamma (t)$ is the thermal noise per unit mass. In the limit of instantaneous collisions, we can assume that the thermal noise has Gaussian statistics, with 
\begin{eqnarray}
    \mathbb{E}[\Gamma (t)] = 0, \quad \mathbb{E}[\Gamma(t)\Gamma(t')] = 2 \gamma \frac{k_BT_b}{m}\delta(t-t'), 
\end{eqnarray}
c.f.~\cite{risken_fokker-planck_1996, van_kampen_chapter_2007, hanggi_stochastic_1982}. The diffusion coefficient is $k_BT_b/m\gamma$. We set the Boltzmann constant, $k_B = 1$, mass, $m = 1$, and friction constant, $\gamma = 1$, to unity. The probability density, $p(x,t)$ to find the particle at time $t$ and coordinate $x$, obeys the Fokker-Planck (FP) equation,
\begin{eqnarray}
\label{eq:FPeq}
    \partial _t p(x,t) &=& \mathcal{L}_{\rm FP}\,p(x,t), 
    \\
    \mathcal{L}_{\rm FP} &\equiv& \partial_x \left[U'(x) + T_b \partial _x\right], 
\end{eqnarray}
where $\mathcal{L}_{\rm FP}$ is the FP operator. We assume that the system is closed, $x \in \mathcal{D} \equiv [x_{\rm min}, x_{\rm max}]$. In this case, the probability is conserved, and we have reflective boundary conditions, which means that the current probability density, $j(x,t)$, defined as $\partial_t p(x,t) = - \partial_x j(x,t)$, is zero at the boundaries, i.e. $j(x_{\rm min},t) = j(x_{\rm max},t) = 0$. The stationary distribution is the Boltzmann distribution, 
\begin{eqnarray}
    \pi ^{T_b}(x) = \frac{1}{Z(T_b)}e^{-\frac{U(x)}{T_b}}, 
\end{eqnarray}
where $Z(T_b) \equiv \int ^{x_{\rm max}} _{x_{\rm min}} \exp[-U(x)/T_b]\,dx$ is the partition function. We assume that the system is initially at thermal equilibrium at temperature $T$,  
\begin{eqnarray}
    p(x,0) = \pi^T(x). 
\end{eqnarray}

\subsubsection{Optimal transport problem for overdamped Langevin dynamics}

The continuity equation,~\EQ{continuityeq}, with the mean local velocity of the process $\rm x(t)$, 
\begin{align}
\label{eq:vel-FP}
    v(x,t) \equiv \frac{j(x,t)}{p(x,t)} =  - U'(x) - T_b \partial _x \ln p(x,t). 
\end{align}
is a FP equation,~\EQ{FPeq}. From~\EQS{Ent-prod-continuous}{vel-FP} the entropy production is explicitly is  
\begin{eqnarray}
    \nonumber
    \Sigma(\tau)&=& \int _\mathcal{D}dx \bigg\{\beta_b U(x) \left[ \pi^T(x) - p(x,t)\right] 
    \\
    \label{eq:SigmaLangevinexp}
    &+& \pi^T(x)\ln \pi^T(x) - p(x,t)\ln p(x,t)\bigg\}. 
\end{eqnarray}

In the case of ~\EQ{FPeq} the minimum of dissipative dynamics can always be achieved with a potential velocity $v$, i.e., a potential force~\cite{benamou_computational_2000,aurell_optimal_2011,aurell_refined_2012,van_vu_thermodynamic_2023}. 

\subsubsection{The Mpemba effect for overdamped Langevin dynamics-- physical interpretation} 

To search for the Mpemba effect we should look at the distance from equilibrium, for example, the KL divergence. Using~\EQS{KLdivdiffSigma}{SigmaLangevinexp}, the KL divergence can be written as 
\begin{eqnarray}
\nonumber
&&D_{\rm KL}\left(p(\tau
    )||\pi^{T_b}\right) = \int_\mathcal{D}dx \bigg\{\beta_b U(x)(p(x,t) - \pi^{T_b}(x)) 
    \\
    &+& p(x,t) \ln p(x,t) - \pi^{T_b}(x) \ln \pi^{T_b}(x)\bigg\}. 
\end{eqnarray}
By restricting our consideration to a range of initial temperatures, one can ask about the instances of minimal KL divergence and minimal total entropy production at times $\tau \gg \max\{t_M, -\lambda_2^{-1}\}$, where $t_M(T)$ is the largest time when a pair $D_{\rm KL}-$curves, from the considered set of initial conditions $\{T_i\}$, cross. 

\begin{figure}
    \includegraphics[width=\columnwidth]{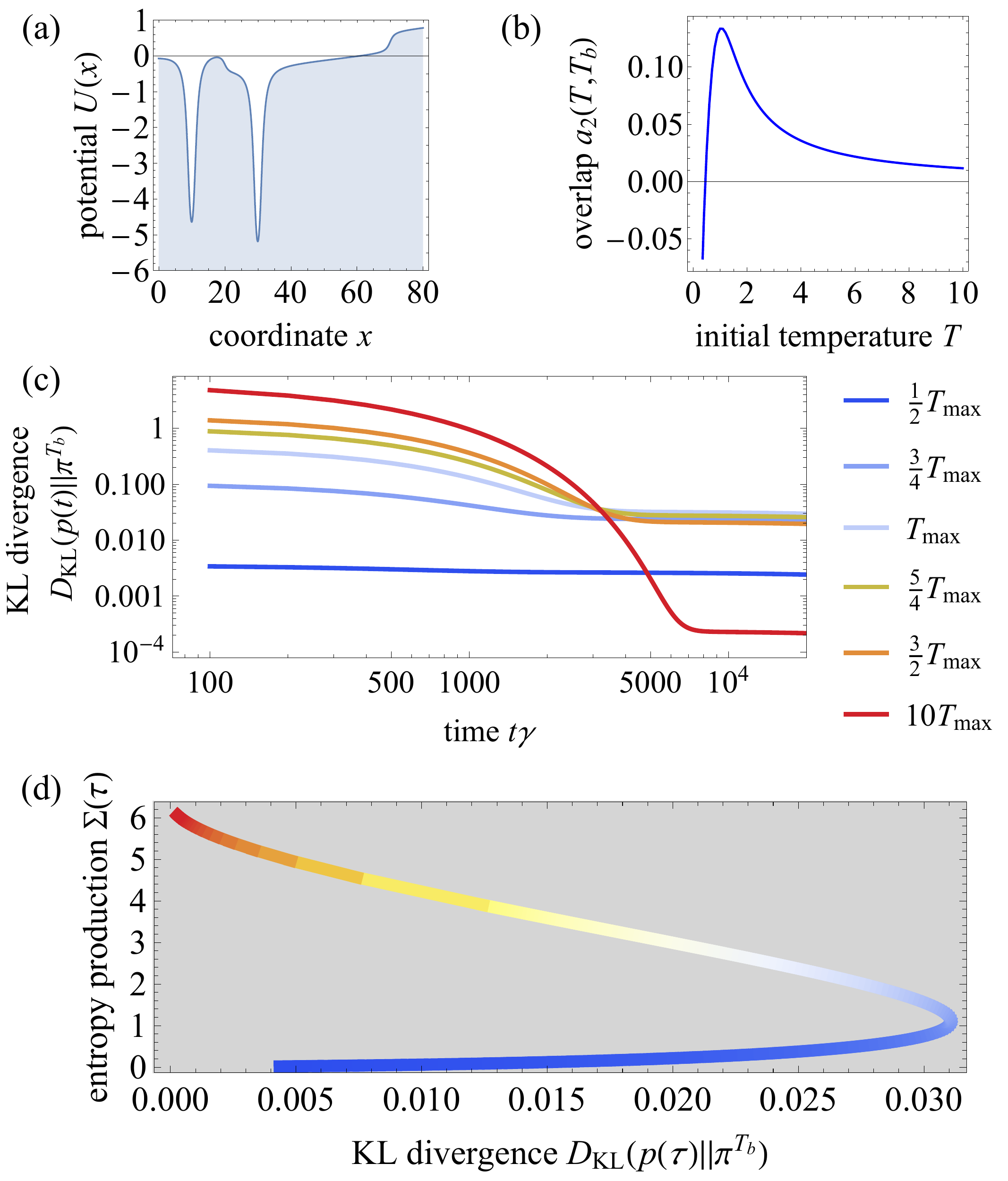}
    \caption{(a) A one dimensional potential, previously considered in~\cite{lu_nonequilibrium_2017} with overdamped-Langevin dynamics,~\EQ{overdampedLangevin} and diffusion constant proportioanl to $T_b = 0.45$. (b) The overlap coefficient $a_2$ is non-monotonic, indicating a Mpemba effect, and has a maximum at $T_{\rm max} = 1.0462$. The eigenvalues are: $\lambda_2 =-2.5\times 10^{-6}$ and $\lambda_3 = - 0.001$. (c) The time dependence of the KL divergence, $D_{\rm KL}(p(t)||\pi^{T_b})$ is computed with approximate $p(t)\approx \pi^{T_b} + a_2 v_2 e^{\lambda_2 t} + a_3 v_3 e^{\lambda_3 t}$. The crossings of the KL divergence curves indicated the Mpemba effect -- for example, the process starting at $10 \,T_{\max}$ (red) by $\tau \gamma = 10^4$ overtakes all shown curves. The process starting at $T_{\max}$ (lightest blue) has maximal KL divergence. Therefore the closest to equilibrium at $\tau \gamma = 10^4$ is the process starting at $10 \, T_{\max}$ (red) and the farthest is one starting at $T_{\max}$ (lightest blue). (d) The parametric plot of the total entropy production $\Sigma(\tau)$ with KL divergence $D_{\rm KL}(p(\tau)||\pi^{T_b})$ at $\tau \gamma = 10^4$. In the interval $T \in \left[T_{\max},10\, T_{\max}\right]$ we have for $T_{\max}$ the minimum of $\Sigma(\tau)$ and maximum of $D_{\rm KL}$, and for $10 \,T_{\max}$ the maximum of $\Sigma(\tau)$ and minimum of $D_{\rm KL}$. Here, within the chosen interval, the optimal transport is at $T_{\rm max}$, but the "fastest" (closest to equilibrium at $\tau$) trajectory has the highest entropy production (at $10\, T_{\max}$).} 
    \label{fig:fig-FP-DKL-entropy-prod-Uarctan-combo.pdf}
\end{figure}

For the example initially introduced in~\cite{lu_nonequilibrium_2017}, we compute KL divergence and the total entropy production. The potential is shown on~\FIG{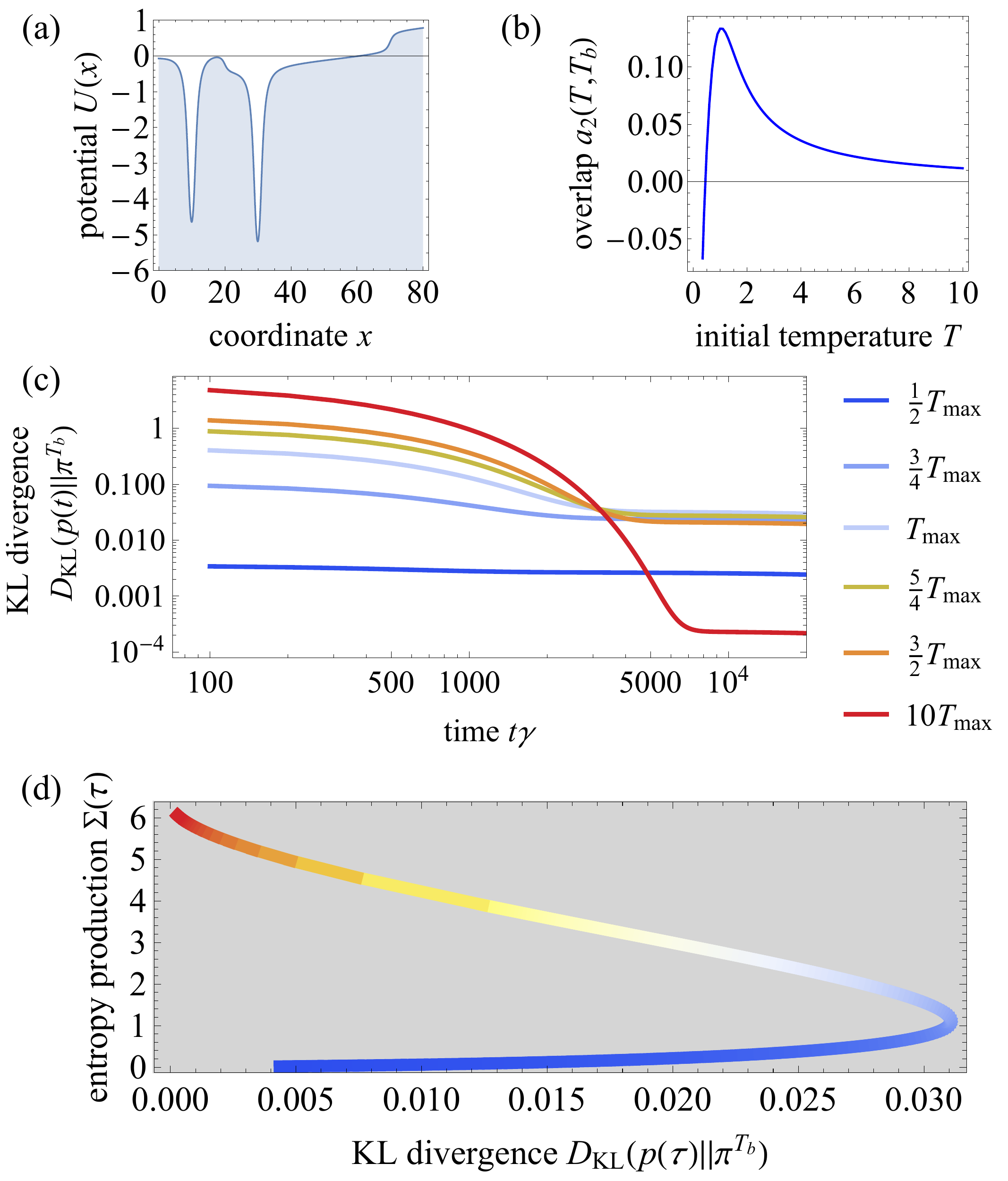}a. The diffusion coefficient was proportional to $T_b = 0.45$. There is a gap between $\lambda_2 = -2.56 \times 10^{-6}$ and $\lambda _3 =-0.001$ and the overlap coefficient $a_2$ is nonmonotonic with a local maximum at $T_{\max} = 1.0462$, indicating a Mpemba effect in cooling (for a range $T > T_b$), see ~\FIG{fig-FP-DKL-entropy-prod-Uarctan-combo.pdf}b. At large time, here $\tau \gamma = 10^4$, we observe that for a range of initial temperatures, here specifically for $T\in [T_{\max},10 \,T_{\max}]$, that the total entropy production is a monotonically decreasing function of the KL divergence -- resulting in minimal KL divergence and maximal total entropy production at $10 \,T_{\max}$ and maximal KL divergence and minimal total entropy production at $T_{\max}$; see~\FIG{fig-FP-DKL-entropy-prod-Uarctan-combo.pdf}d. Hence the optimal transport in time $\tau$ for a range of initial temperatures $T\in [T_{\max},10 \, T_{\max}]$ happens for $T = T_{\max}$, but this is also the "slowest" trajectory, as it is farthest from equilibrium at the chosen time $\tau\gamma = 10^4$. While the "fastest" trajectory, the closest to equilibrium at $\tau$, among those labeled with initial conditions from $T\in [T_{\max},10 \, T_{\max}]$ is the one starting at $10 \,T_{\max}$, at the same time this trajectory also has the highest total entropy production of the set. To summarize, the above example shows a case of an often "antipodal" relation between the "optimal" transport (minimal total entropy production) and "fast" relaxation (here, the Mpemba effect). 

The optimal transport problem is typically defined with a well-defined starting point $p^A$ and well-defined end $p^B$ after a finite time $\tau$. The optimal transport is the one that minimizes the total entropy production by altering the dynamics with specified control parameters. Above, we did not change the dynamics; instead, we considered a range of initial conditions, and we asked which initial condition minimizes the total entropy production and, after a large but finite time $\tau$, how far away from the equilibrium distribution is the probability distribution at time $\tau$. 

Next, suppose we vary the potential in a continuous manner with a time-dependent control parameter, $\delta(t)$, and let us assume the potential variations are with fixed temporal endpoints, $U[\delta(t=0)] = U[\delta(t = \infty)]$. Also, suppose that among different variations of $\delta (t)$, there is a protocol, $\delta _{\rm SM}(t)$, such that there is no overlap to the slowest mode, i.e., for that protocol, $a_2= 0$, and the KL divergence is minimal,
\begin{eqnarray}
    \min _{\delta(t)} D_{\rm KL}\left(p(\tau
    )||\pi^{T_b}\right). 
\end{eqnarray}
Here $\tau$ is a sufficiently large time, meaning $\tau \gg \max\{t_M, -\lambda_2^{-1}\}$, where $t_M(T)$ is the largest time when a pair $D_{\rm KL}-$curves, from the considered set of protocols conditions $\{\delta(t)\}$, cross. The KL divergence at $\tau$ is the difference between the entropy production at infinity and at $\tau$,~\EQ{KLdivdiffSigma}. Since the entropy production at infinity only depends on the initial condition and the equilibrium, see~\EQ{SigmaLangevinexp}, the minimum KL divergence corresponds to maximal total entropy production. 
Thus, for potential variations with fixed temporal endpoints, the Strong Mpemba effect and optimal transport generically would not happen for the same dynamics. 

The question we ask here is whether the same intuition will hold in the discrete case -- i.e., if there are cases where the dynamics corresponding to the Mpemba-like phenomena between the source and the target distribution are also optimal. 

\subsection{Three-level system}

We consider a fully connected three-level system with energies $\{E_1, E_2, E_3\}$. The Mpemba effect on such systems was already considered in~\cite{lu_nonequilibrium_2017} and recently as a function of dynamics in~\cite{bera_effect_2023}. We define the clockwise direction as $1 \to 2 \to 3 \to 1$ and clockwise the transition rates are 
\begin{eqnarray}
\label{eq:ratesRmat}
   &&R_{21} = \gamma e^{-\frac{1}{2}\beta_b(E_{2} -E_{1})},\, 
   R_{13} = \gamma e^{-\frac{1}{2}\beta_b(E_{1} -E_{3})}, 
   \\
   \label{eq:ratesRmatdelta}
    &&R_{32}(\delta) = \gamma e^{-\beta_b(E_{3} -E_{2})\delta}, 
\end{eqnarray} 
where $\gamma^{-1} = 1$ sets the unit of time, and $R_{32}$ has an additional control parameter, $\delta \in [0,1]$, for its magnitude. DB, \EQ{detailedbalance}, sets the corresponding "counter-clockwise" transitions. DB does not prescribe the dynamics; it just sets the ratio between the forward and backward rates. By changing $\delta$, we change the magnitude of the rates between states $2$ and $3$ -- because of DB, this local change affects all of the currents $j_{xy}$. While in a larger graph, only currents connected to the two nodes involved are affected. The parameter $\delta$ is often called the load factor, and it has been studied in the context of molecular motors~\cite{kolomeisky_molecular_2007,lau_nonequilibrium_2007,kolomeisky_motor_2013}, differential mobility~\cite{teza_rate_2020}, Markov jump processes~\cite{remlein_optimality_2021}, and recently by the authors, in the context of anomalous thermal relaxations~\cite{bera_effect_2023}. The conservation of probability sets the diagonal elements -- the columns of the $R$ matrix sum to zero, i.e. 
\begin{eqnarray}
\label{eq:rateRmatdiag}
    R_{xx} &=& - \sum_{\substack{y\in \Omega\\y \neq x}} R_{yx},\quad \forall x \in \Omega. 
\end{eqnarray}
In general, the rate matrix depends on the properties of the system, the environment, and time. However, we restrict our considerations below to rate matrices that depend solely on the bath temperature, $T_b$, and a control parameter specifying the dynamics, which we introduce below. Lastly, note that the three-level system considered here is fully connected. Thus, the Wasserstein distance is equal to the total variation distance, see~\EQ{Wasserstein-totalvariation}. 

We look at a situation where for the given initial temperature $T$ and bath temperature $T_b$, there is a Strong Mpemba effect at load factors $0 <\delta < 1$. 
One such case is on~\FIG{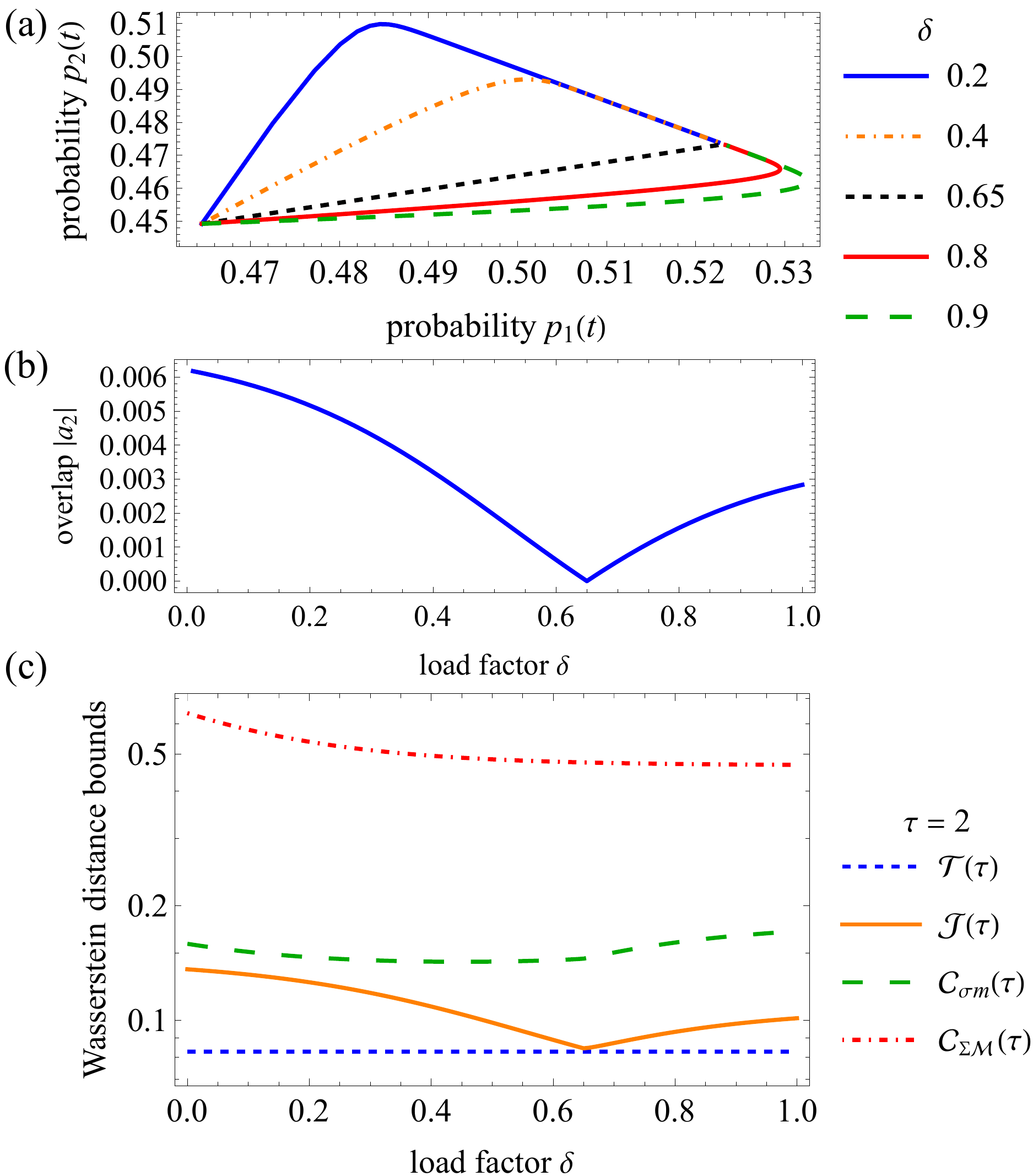}. 
\begin{figure}
    \includegraphics[width=\columnwidth]{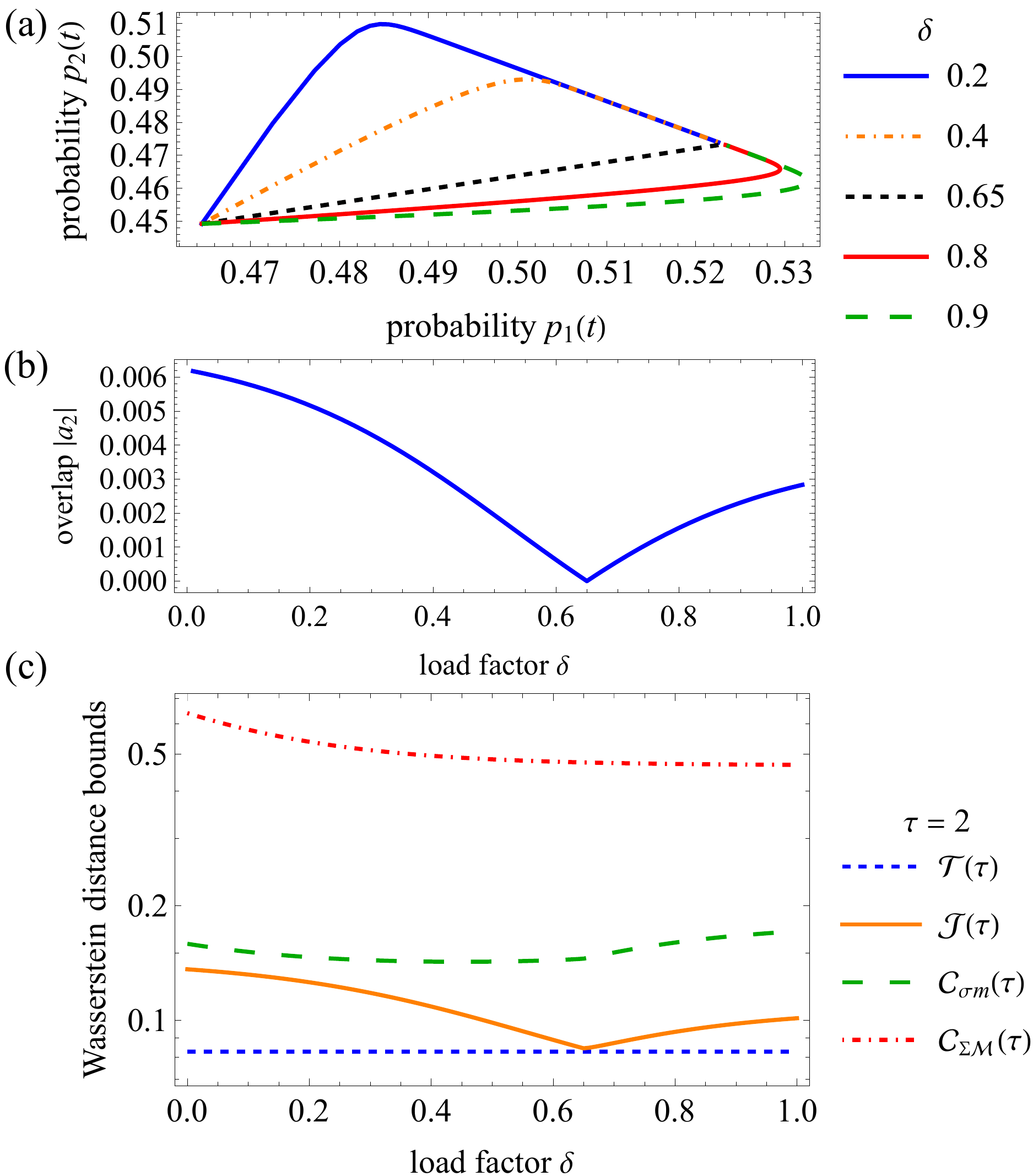}
    \caption{(a) Relaxation trajectories of probabilities of state occurrences, $p_1(t)$ and $p_2(t)$, for a three-level system. The energies of the system are $\{0,0.1\,T_b,5\,T_b\}$, the initial temperature is $T = 2.97\,T_b$ and $\tau = 2/\gamma$ (with $T_b = 1$ and $\gamma = 1$). Different colors represent different dynamics, parameterized by the load factor $\delta$, introduced in~\EQ{ratesRmatdelta}. The dashed black line is for $\delta_{\rm SM} = 0.65$, which corresponds to the Strong Mpemba effect. The relaxation with $\delta_{\rm SM}$ is along a straight line, as the projection on the second eigenvector is zero. (b) The overlap $|a_2|$ vanishes for $\delta_{\rm SM} = 0.65$, which corresponds to the Strong Mpemba effect. (c) Total variation distance $\mathcal{T}(\tau)$, flow cost $\mathcal{J}(\tau)$ and upper bounds $\mathcal{C}_{\sigma m}(\tau)$ and $\mathcal{C}_{\Sigma \mathcal{M}}(\tau)$ as a function of the load factor $\delta$. The optimal transport has a minimal flow cost. In this example, the minimal flow cost and the Strong Mpemba effect happen at the same load factor, $\delta = \delta_{\rm SM}$.}
    \label{fig:2023-07-10-SM_Tb_1T_2.9697delta_0.65eps_0.1E3_5.tau_2-figs.pdf}
\end{figure}
We observe that in the cases where the gap between $\lambda_2$ and $\lambda_3$ is large, $(\lambda_2 - \lambda_3)/\tau \ll 1$, the flow cost $\mathcal{J}(\tau)$ has a minimum at the same load factor as the overlap coefficient $|a_2|$, indicating that in that case, the optimal transport is the one where the Strong Mpemba occurs. The dynamic state mobility $m(\tau)$ saturates at large times because it is proportional to the difference in activities,~\EQ{Onsager-like-coefs}. Thus the even later times' contributions to the bounds $\mathcal{C}_{\sigma m}(\tau)$ and $\mathcal{C}_{\Sigma \mathcal{M}}(\tau)$ are mainly entropic -- due to the entropy production rate and the entropy production. 

For smaller gaps or shorter times, the contribution of the fast mode also matters, and the entropy production rate is not minimal for the same load factor as the occurrence of the Strong Mpemba effect. Looking at larger times $\tau$, in this case, does make things more entropy-dominated and slow-mode-dominated. Mobility likewise again saturates in finite time. 

Similarly, for the fully-connected four-state system, with a large gap between the two slowest modes and at times larger than the slowest mode, the minimum of the flow cost and Strong Mpemba effect happen at the same load factor $\delta$. Note that in the case of the four-state system, things are more complicated as the eigenvalues of the rate matrix can cross~\cite{teza_eigenvalue_2023}. 

The main result here is that in the Markov jump processes case, the Strong Mpemba effect and the optimal transport can sometimes occur for the same dynamics. We show that a small tweak in the dynamics could be used to minimize entropy dissipation without sacrificing mobility. These results are in stark contrast with the intuition gained from the continuous case of overdamped-Langevin dynamics. 

\section{Discussion} 

In discrete systems, a small change in the long time limit of the dynamic state mobility might influence a large change in the total entropy dissipation, while in the overdamped Langevin case, the long time limit of the dynamic state mobility is constant with respect to the considered dynamics changes, and proportional to the diffusion constant, $T_b$. The two cases also differ in the allowable probability currents -- in the continuous cases considered, the probability currents are continuous, while in the discrete case, we can have quite a wide distribution of currents restricted only by detailed balance.   

We find seemingly counter-intuitive cases in which the Strong Mpemba effect and the minimal Wasserstein distance occur at the same load factor. The exponentially faster relaxation to thermal equilibrium also occurs with minimal entropy production for specific types of dynamics. We argue that such a scenario is highly surprising, especially considering our continuous paradigm -- the overdamped Langevin dynamics with continuous variations of continuous potential, where the Strong Mpemba effect is generally observed together with a high entropy production. 

More work is needed to verify our findings in a discrete case for a larger reaction network and to specify what kind of variations of the dynamics are needed to observe the Mpemba effect and the optimal transport for the same protocol. Another consideration is the size of the system and the relative size of the perturbation of the dynamics needed to have the optimal transport and Mpemba effect happen for the same protocol. 

Somewhat conceptually related to our results are the results of elastic network alterations where even small local perturbations in specific networks can change their macroscopic responses, flow, and functionality of these networks, see, e.g., and references within~\cite{rocks_hidden_2021}. 
 
Like optimal transport, the Mpemba effect could be helpful in designing efficient samplers, optimal heating and cooling protocols, and preparations of state. Real-world applications of our findings will depend on the feasibility of altering the dynamics in such a way as to have both the Mpemba effect and optimal transport.

\begin{acknowledgments}
MV, SB, and MW acknowledge discussions with Gregory Falkovich, Gianluca Teza, Amartyajyoti Saha, and Aaron Winn. This material is based upon work supported by the National Science Foundation under Grant No.~DMR-1944539.
\end{acknowledgments}

\bibliographystyle{apsrev4-1}
\bibliography{references.bib}
\end{document}